\documentclass[preprint,5p]{elsarticle}
\usepackage{graphicx}
\usepackage{dcolumn}
\usepackage{bm}
\usepackage{siunitx}
\usepackage{color}
\usepackage{amsmath,amssymb,amsthm,color}

\begin{document}

\title{IR fixed points in $SU(3)$ gauge Theories}


\author{K.-I. Ishikawa}
\address{Graduate School of Science, Hiroshima University,Higashi-Hiroshima, Hiroshima 739-8526, Japan}

\author{Y. Iwasaki}
\address{Center for Computational Sciences, University of Tsukuba,Tsukuba, Ibaraki 305-8577, Japan}

\author{Yu Nakayama}
\address{ Walter Burke Institute for Theoretical Physics,California Institute of Technology,  Pasadena, CA 91125, USA}

\author{T. Yoshie}
\address{Center for Computational Sciences, University of Tsukuba,Tsukuba, Ibaraki 305-8577, Japan}

\date{\today}

\begin{abstract}
We propose a novel RG method to specify the location of the IR fixed point in lattice gauge theories and apply it 
to the $SU(3)$ gauge theories with $N_f$  fundamental fermions.
It is based on the scaling behavior of the propagator through the RG analysis with a finite IR cut-off, which we cannot remove in the conformal field theories in sharp contrast with the confining theories.
The method also enables us to estimate the anomalous mass dimension in the continuum limit at the IR fixed point. We perform the program for $N_f=16, 12,  8 $ and $N_f=7$ and indeed identify the location of the IR fixed points in all cases.

\end{abstract}

\maketitle

Scale invariance, or more precisely conformal invariance  has become a fundamental concept in understanding the universal aspects of the nature from the Planck scale to the Hubble scale. They appear not only in critical phenomena of condensed matter physics, but also in quantum gravity, high energy particle phenomenology, and all the way up to cosmology \cite{Nakayama:2013is}. Many conformal field theories, however, are strongly coupled, and much remains unsolved in their theoretical understanding. In particular, when realized by gauge theories, the constructive approaches to the conformal fixed points are still rudimentary \cite{review}. The aim of this article is to clarify some important aspects of these constructive approaches and offer one simple criterion on conformal invariance.

Obviously, the central question is to locate the IR fixed point within a given class of theories. In this article we propose a novel and simple RG method to specify the location of the IR fixed point in lattice gauge theories by studying the scaling behavior of the propagator. 
We will apply the technique to the $SU(3)$ gauge theories with $N_f$  fundamental fermions (within the conformal window), and estimate the anomalous mass dimension.
We perform this program for $N_f=16, 12,  8 $ and $N_f=7$,
and indeed identify the location of the IR fixed points in all cases.

We constructively define gauge theories on Euclidean plane $\mathbf{R}^4$ as the continuum limit of lattice gauge theories on the Euclidean lattice of the size $N_x=N_y=N_z=N$ and $N_t=r N$ ($r$ being an aspect ratio, which is fixed as $r=4$ throughout the article),  
taking the limit of the lattice space $a \rightarrow 0$ and  $N \rightarrow \infty$, with $L =N \, a$  and $L_t =N_t \, a$ fixed.
When  $L$ and/or $L_t$ are finite, the system is bounded by an IR cutoff $\Lambda_{\mathrm{IR}} \sim 1/L$.
We impose an anti-periodic boundary condition in the time direction for fermion fields and
periodic boundary conditions otherwise.
In conformal field theories the IR cutoff is an indispensable ingredient because there is no other natural scale to compare, which will be further elucidated in this article.


Our general argument that follows can be applied to any gauge theories with fermions in arbitrary (vector-like) representations, but to be specific, we focus on $SU(3)$ gauge theories with $N_f$ fundamental Dirac fermions. For the lattice regularization of the action, we employ the Wilson quark action and the RG improved gauge action\cite{RG action} (also known as the Iwasaki gauge action in the literature).

Given the regularized action, the theory is defined by two parameters; the bare coupling constant $g_0$ and the bare degenerate quark mass $m_0$ at ultraviolet (UV) cutoff.
We also use, instead of $g_0$ and $m_0$, 
$\beta={6}/{g_0^2}$
and the hopping parameter
$K= 1/2(m_0a+4)$. 

As for observables, together with the plaquette and the Polyakov loop in each space-time direction, we measure the quark mass $m_q$ defined 
through Ward-Takahashi identities
\begin{equation}
m_q 
=  \frac{\langle 0 | \nabla_4 A_4 | {\rm PS} \rangle}
        {2\langle  0 | P | {\rm PS} \rangle}, 
\label{eq:quark-mass}
\end{equation}
where $P$ is the pseudo-scalar density and $A_4$ the fourth component of the
local axial vector current, with renormalization constants being suppressed.
The quark mass $m_q$ defined in this way only depends on $\beta$ and $K$ up to order $1/N$ corrections.


One of the most important observables we will study is the $t$ dependence of the propagator of the local meson operator in the $H$ channel:
\begin{equation}
G_H(t) = \sum_{x} \langle \bar{\psi}\gamma_H \psi(x,t) \bar{\psi} \gamma_H \psi(0) \rangle \ ,
\label{propagator}
\end{equation}
where the summation is over all the spatial lattice points.
In this paper, we mostly focus on the pseudo-scalar (PS) channel $H=PS$, and the subscript $H$ is suppressed hereafter.

\begin{figure*}[htb]
\begin{center}
			\includegraphics [width=7.5cm]{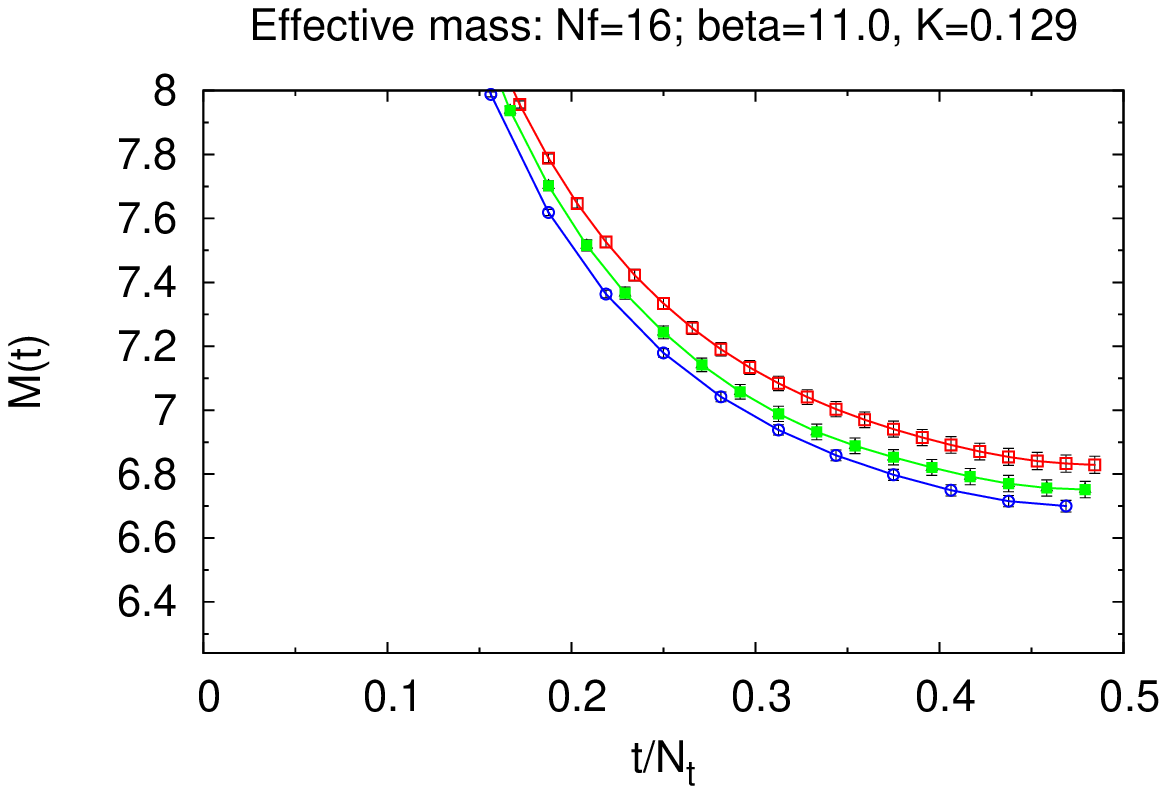}
	\hspace{0.5cm}
	\includegraphics [width=7.5cm]{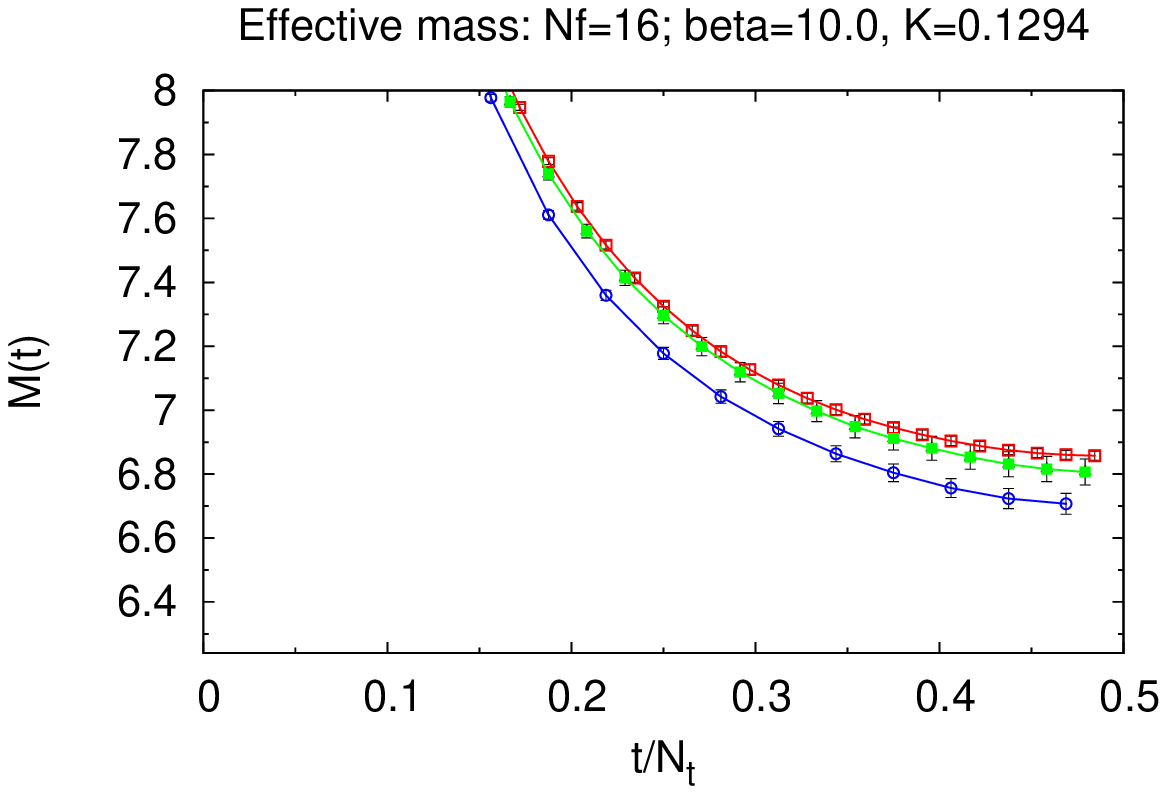}
\vspace{1cm}
\caption{Scaled effective mass plots for $N_f=16$ at $\beta=11.0$ and $10.0$: 
three sets of symbols are $N=16$ (red square), $N=12$ (green circle), $N = 8$ (blue triangle).}
\label{effective mass plot}
\end{center}
\end{figure*}

%
%

\begin{figure*}[htb]
\begin{center}
\includegraphics [width=7.5cm]{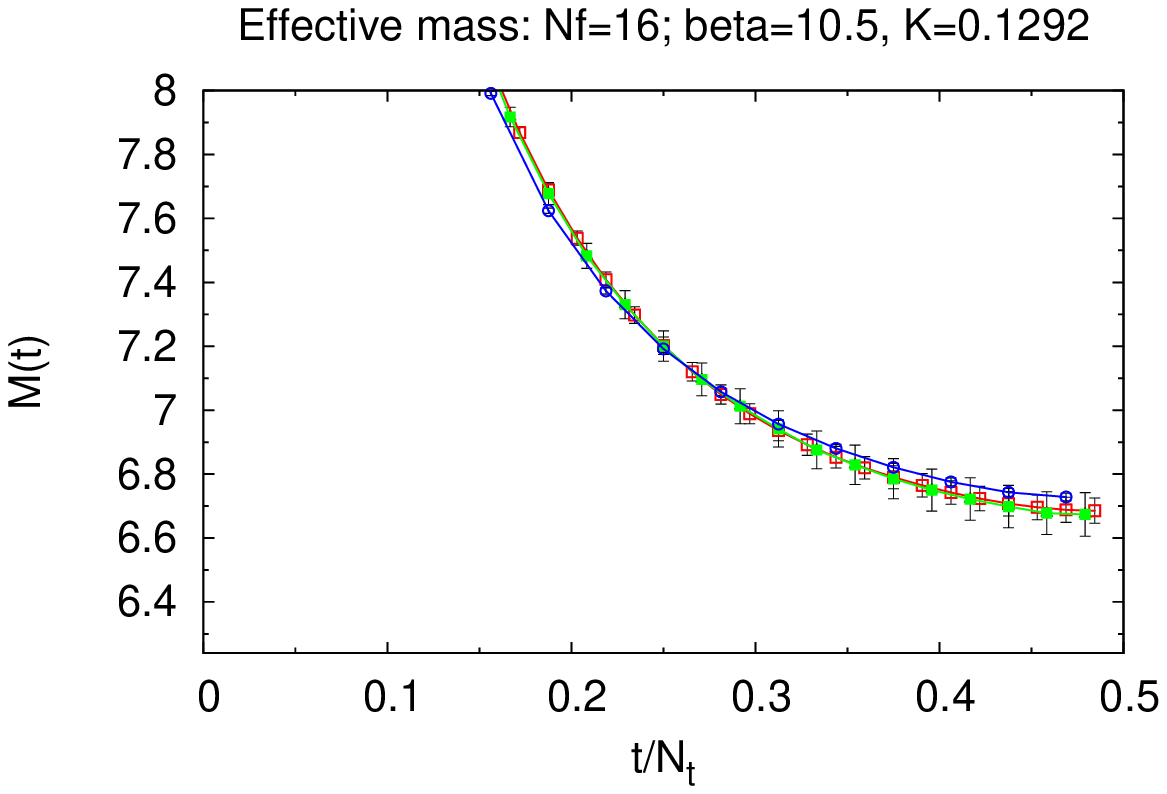}
\hspace{1cm}
	\includegraphics [width=7.5cm]{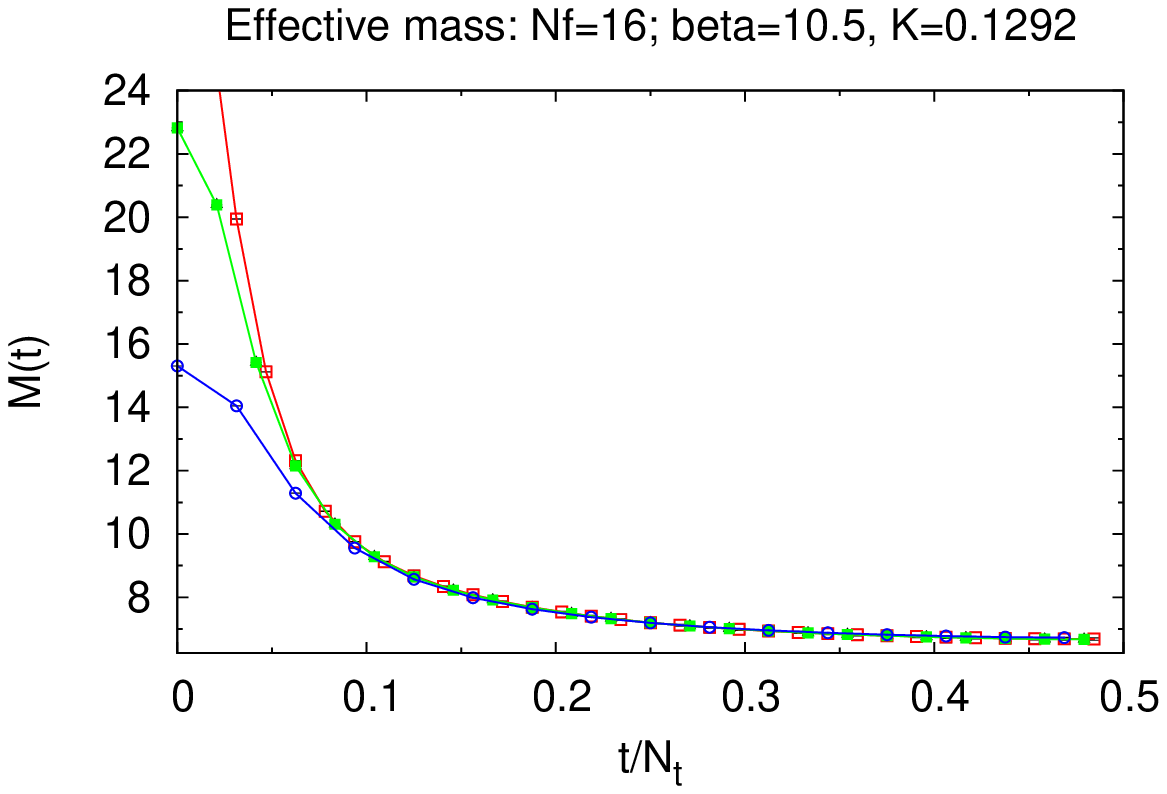}
\vspace{1cm}
\caption{Scaled effective mass plots for $N_f=16$ at $\beta=10.5$: the left panel is an enlarged one of the right panel;
three sets of symbols are $N=16$ (red square), $N=12$ (green circle), $N = 8$ (blue triangle).}
\label{nf16}
\end{center}
\end{figure*}

In order to investigate the large $t$ behavior of a propagator,
we define the effective mass $m(t)$ through  
\begin{equation}\frac{\cosh(m(t)(t-N_t/2))}{\cosh(m(t)(t+1-N_t/2))}=\frac{G(t)}{G(t+1)}.\label{effective mass}\end{equation}
When boundary effects can be neglected, it reduces to
\begin{equation}m(t) = \ln \frac{G(t)}{G(t+1)}.\label{simple effective mass}\end{equation}
In the case of exponential-type decay the effective mass approaches a constant value 
in the large $t$ regime, which we call a plateau.

Before non-perturbative discussion, let us first recall the perturbative result.
Within the two-loop  perturbation theory, the RG beta function for the $SU(3)$ gauge coupling constant is given as
\begin{align}
\mathcal{B}(g) = -\frac{(33-2N_f)}{48\pi^2} g^3 -\frac{ \left(102 - \frac{38}{3}N_f\right) }{(16\pi^2)^2}g^5 + \mathcal{O}(g^7) \ . \label{betaff}
\end{align}

The fixed point $\mathcal{B} (g^*) = 0$ exists for $8.05 < N_f <16.5$ within the two-loop approximation\cite{Banks1982}. 
When $N_f=16$, the IR fixed point is located at
$\beta=11.48.$
Since this coupling constant is small, we may trust the perturbative computations for $N_f=16$.
We will compare the non-perturbative calculation with this value.

When $N_f$ decreases,  $g^*$ increases, at least in the perturbation theory, and therefore non-perturbative effects become important. 
The smallest $N_f$ where the fixed point exists is denoted as $N_f^c$ and the range of flavors $N_f^c \le N_f  \le 16$ is called the ``conformal window''.
The lower bound of the conformal window can only be determined non-perturbatively.
Our earlier studies \cite{iwa2004}\cite{coll2014}  strongly suggest the conjecture that the conformal window is $7 \le N_f \le 16.$
However, the conjecture is based on indirect logics.
In this article, we will present more direct evidence supporting the conjecture.



%

Let us study the RG properties of the propagator in the vicinity of the fixed point.  
The RG equation for the RG transformation induced by the change of the UV renormalization scale $\mu'=\mu/s$, followed by a space-time scale change by a factor $1/s$ (see e.g. \cite{DelDebbio:2010ze}), relates the propagator with different parameters as 
\begin{equation}G(t; g, m_q, N, \mu)={\left(\frac{N'}{N}\right)}^{3-2\gamma} G(t{'}; g{'}, m'_q, N{'}, \mu )\label{mass dimension 2} .\end{equation}
Here $N{'}=N/s$ and $t{'}=t/s.$
The relation between $g'$ and $g$ and $m'_q$ and $m_q$ are determined by the beta function $\mathcal{B}$ and the mass anomalous dimensions $\gamma$.

The UV renormalization scale $\mu$ in lattice theories is set by the inverse lattice spacing $a^{-1}.$  In Eq. \eqref{mass dimension 2} the parameter $\mu$ is common on both sides and therefore $\mu$ may be suppressed in the notation, but we should keep in mind that the RG equation \eqref{mass dimension 2} implicitly assumes the continuum limit $N, N' \to \infty$ to avoid the effect of the UV cutoff.


Let us first discuss the case in which we are at the fixed point, i.e. $g' = g = g^*$ and $m_q' = m_q= 0$ so that $\mathcal{B}=0$ and $\gamma = \gamma^*$. 
In this case, the propagator may have  simplified notation as
\begin{equation} \tilde{G}(\tau,N)= G(t,N).\end{equation}
with $\tau =t/N_t$. The variable $t$ takes $0, 1, 2, \cdots, N_t-1$ so that $0 \leq \tau \leq 1$.   
In terms of $\tau$, the RG relation eq.(\ref{mass dimension 2}) reduces to
\begin{equation}\tilde{G}(\tau; N)={\left(\frac{N^{'}}{N}\right)}^{3-2\gamma^*} \tilde{G}(\tau; N^{'})\label{mass dimension} \ . \end{equation}
Strictly speaking, this equation is satisfied in the limit $N, N'\rightarrow \infty$.



To state our proposal concretely, we define the scaled effective mass  $\mathfrak{m}(t;N)$ as
\begin{equation}\mathfrak{m}(t, N) = N \ln \frac{G(t, N)}{G(t+1, N)}.\label{scaled effective mass}\end{equation}
In the continuum limit $N \rightarrow \infty$ Eq.\ (\ref{scaled effective mass}) reduces to the form
\begin{equation}\mathfrak{m}(\tau ,N)= - \partial_\tau \ln \tilde{G}(\tau, N)\label{effective mass relation}\end{equation}
The crucial observation, which will be the core of our proposal is that, combining Eqs.(\ref{mass dimension}) and (\ref{effective mass relation}), the scaled effective mass does not depend on $N$ as a function of $\tau$: 
 \begin{equation}
\mathfrak{m}(\tau, N)=\mathfrak{m}(\tau, N^{'})\label{equality of scaled effective mass}\end{equation}
at the fixed point. Therefore, the agreement of the scaled effective mass as a function of $N$ and $\tau$ are stringent tests of the fixed point.

Suppose that we are away from the fixed point (i.e $g \neq g^*$ while $m_q = 0$) in contrast. The scaled effective mass in the vicinity of the fixed point would instead show the following behavior in the leading term
\begin{align}
\label{off-fixed-point}
\mathfrak{m}(\tau,g,N) = \mathfrak{m}(\tau,g,N') + \mathcal{B}(g) \ln\left(\frac{N'}{N}\right) \partial_g \mathfrak{m}(\tau,g,N')\ . 
\end{align}

We can rewrite Eq. \ref{off-fixed-point} in a more suggestive form
\begin{align}
\mathcal{B}(g)= \frac{\mathfrak{m}(\tau,g,N)-\mathfrak{m}(\tau,g,N')}{\ln\left(\frac{N'}{N}\right) \partial_g  \mathfrak{m}(\tau,g,N')},
\label{beta_function}
\end{align}
which may be used to compute $\mathcal{B}(g)$  in the vicinity of the fixed points once we numerically compute $\mathfrak{m}(\tau,g,N)$.
This formula is satisfied in the limit $N, N'\rightarrow \infty$.

Our strategy is as follows. With given $N_f$ and $\beta$, we tune the quark mass (defined through Ward-Takahashi identity) to be zero. Then we numerically compute the meson propagator on the lattice. For each choice of the lattice size $N$, we
plot the effective mass defined by Eq. (\ref{effective mass}) in terms of the scaled time $\tau$. As we explained, generically, the scaled effective mass do not coincide with each other as a function of $\tau$ at a given value of $\beta$ but different values of $N$. However, if we find the fixed point value $\beta$, the plots for different $N$ must coincide with each other.

In this article, we 
perform numerical simulations   on the three lattices with size $8^3\times 32, 12^3\times 48$ and $16^3\times 64$  with the aspect ratio of $r = 4$.
Note that although the explicit form of $\mathfrak{m}(\tau,N)$ depends on $r$, the relation Eq. (\ref{equality of scaled effective mass}) itself is valid for any $r$.

Let us show in Fig. 1 the scaled effective mass plots in two cases of many such examples: we take $N_f=16$ at $\beta=11.0$ and $\beta=10.0.$ The asymptotic behaviors of three sets of data points and the lines connecting them on $N=8 ,12$ and $16$ lattices, do not coincide with each other. We may conclude that these values of $\beta$ do not correspond to the fixed point. On the other hand, as we will see in  Fig  2, if we take $\beta =10.5$, then the three plots and the lines do coincide within the standard deviation. 
Based on the RG relations, we claim that this is the value of the gauge coupling constant at the fixed point.

We perform this program for $N_f=7, 8, 12, 16$  
on lattices with size $8^3\times 32, 12^3\times 48$ and $16^3\times 64$. By narrowing down the region where the scaled effective mass 
$\mathfrak{m}(\tau,N)$  becomes close for different $N$, we identify the point $\beta^*$ where they agree with each other
 within one standard deviation. 
Note that the scaled effective mass as a function of $\tau$ in general does depend on $N$,  so the coincidence of the {\it three curves} only at a particular common value of $\beta$ is dramatic. We eventually find that this occurs for all the cases we study.
 



\begin{figure*}[htb]
\begin{center}
		\includegraphics [width=7.5cm]{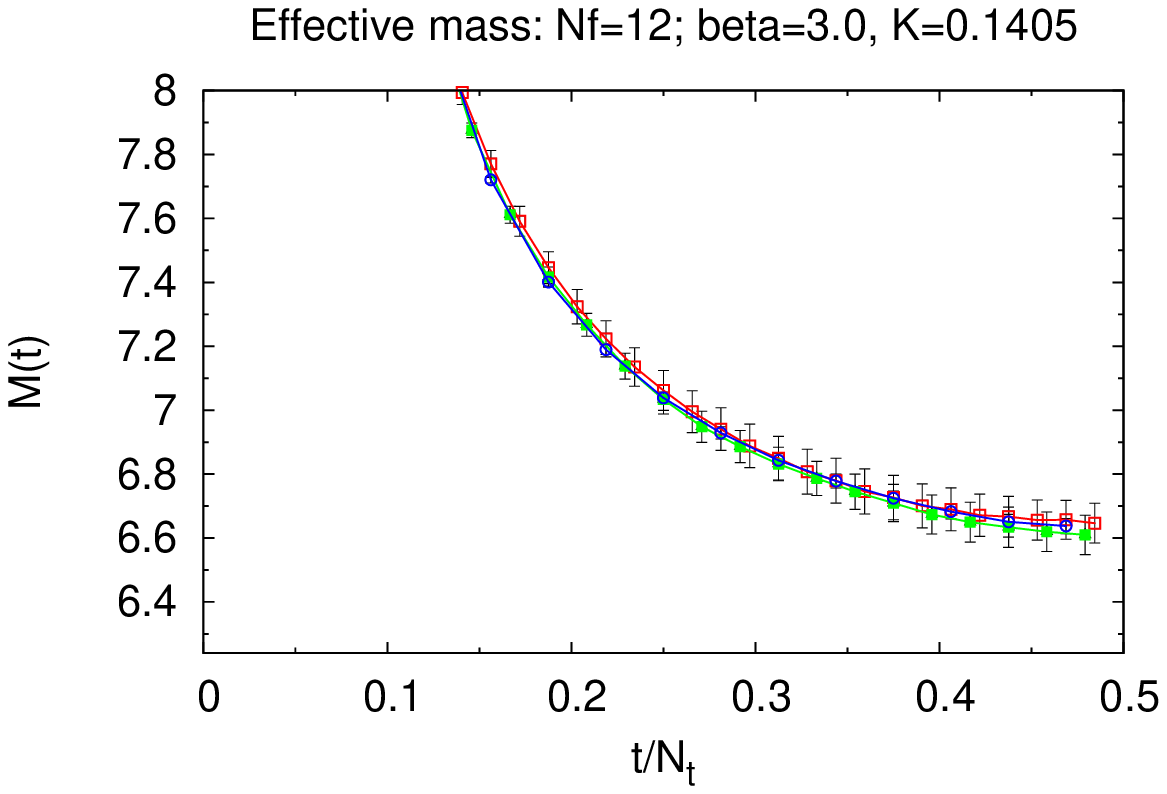}
\hspace{0.5cm}
			\includegraphics [width=7.5cm]{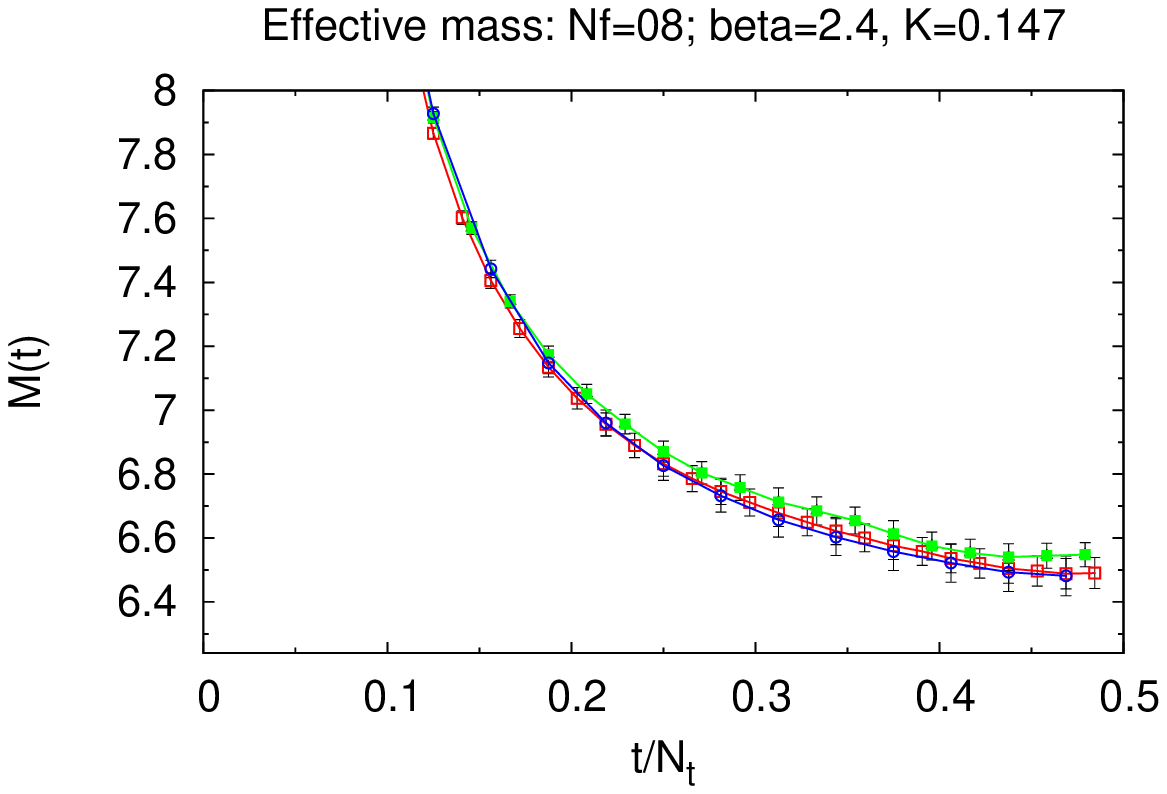}
\vspace{1cm}
\caption{Scaled effective mass plots for $N_f=12$ at $\beta =3.0$ and $N_f=8$ for  $\beta=2.4$;
three sets of symbols are $N=16$ (red square), $N=12$ (green circle) and $N = 8$ (blue triangle).}
\label{effective mass plot; 12 and 8}
\end{center}
\end{figure*}

\begin{table}
\caption{Simulation parameters for the cases we identify the IR fixed points:
the first column $N$ is the lattice size, the second column $N_{\mathrm{traj}}$ is the number of trajectories, the third "acc" is the acceptance ratio for the global Metropolis test, "plaq" is the value of the plaquette and $m_q$ is the  quark mass defined by eq.(\ref{eq:quark-mass}).}
\label{Simulation parameter}
\sisetup{table-alignment=center}
\begin{tabular}{lcSSS}
\hline
\hline
\multicolumn{5}{c}{{$N_f=16$, $\beta=10.5$, $K=0.1292$}}\\
\hline
{$N$} & {$N_{\mathrm{traj}}$} & {{acc}} &  {{plaq}} & {{$m_q$}}\\
\hline
16 & 2000 & \num{0.59(1)} & \num{0.92255(1)} & \num{-0.0063(1)} \\
12 & 4000 & \num{0.77(1)} & \num{0.92255(1)} & \num{-0.0053(1)} \\	
08 & 4000 & \num{0.89(1)} & \num{0.92257(1)} & \num{0.0003(5)} \\
\hline
\hline
\multicolumn{5}{c}{{$N_f=12$, $\beta=3.0$, $K=0.1405$}}\\
\hline
{$N$} & {$N_{\mathrm{traj}}$} & {{acc}} &  {{plaq}} & {{$m_q$}}\\
\hline
16 & 3000 & \num{0.68(1)} & \num{0.74416(2)} & \num{-0.002(1)} \\
12 & 3000 & \num{0.84(1)} & \num{0.74415(1)} & \num{-0.002(1)} \\	
08 & 4000 & \num{0.94(1)} & \num{0.74419(2)} & \num{ 0.004(1)} \\
\hline
\hline
\multicolumn{5}{c}{{$N_f=8$, $\beta=2.4$, $K=0.147$}}\\
\hline
{$N$} & {$N_{\mathrm{traj}}$} & {{acc}} &  {{plaq}} & {{$m_q$}}\\
\hline
16 & 4000 & \num{0.72(1)} & \num{0.67620(1)} & \num{-0.007(1)}  \\	
12 & 4000 & \num{0.84(1)} & \num{0.67620(1)} & \num{-0.006(3)}  \\	
08 & 3000 & \num{0.93(1)} & \num{0.67622(2)} & \num{-0.0005(5)} \\
\hline
\hline
\multicolumn{5}{c}{{$N_f=7$, $\beta=2.3$, $K=0.14877$}}\\
\hline
{$N$} & {$N_{\mathrm{traj}}$} & {{acc}} &  {{plaq}} & {{$m_q$}}\\
\hline
16 & 4000 & \num{0.72(1)} & \num{0.65931(1)} & \num{-0.0017(2)} \\
12 & 4000 & \num{0.85(1)} & \num{0.65931(1)} & \num{-0.0005(3)} \\
08 & 5000 & \num{0.94(1)} & \num{0.65941(3)} & \num{ 0.0047(6)} \\
\hline
\hline
\label{tab:nf16}
\end{tabular}
\end{table}

The algorithms we employ are the blocked HMC algorithm \cite{Hayakawa:2010gm} in the case $N_f=2\, \mathbb{N}$ and the RHMC algorithm \cite{Clark:2006fx} for $N_f=1$ in the case $N_f=2\, \mathbb{N} +1$.
We show the parameters of simulations for the cases we identify the IR fixed points in Table I.
The masses of quarks are expressed in units of $a^{-1}$.
We choose the run-parameters in such a way that the acceptance of the global Metropolis test is about $60\%\sim 90\%.$
The statistics are 1,000 MD trajectories for thermalization and $1000\sim4000$ MD trajectories for the measurement.
We estimate the errors by the jack-knife method
with a bin size corresponding to 100 HMC trajectories.

Before examining our numerical results, we have a few comments in order.

Firstly, for gauge configuration generation we have to be very careful to choose the lowest energy state,
not quasi-stable states in the conformal region.
As stressed in Ref.\cite{coll2014}, there are quasi-stable states, which persist for long time for a HMC algorithm.

Secondly, in contrast with the confining phase,  when the system is either in the deconfining phase at high temperature or in the conformal region, it is not hard to perform simulations at zero quark mass. It is even possible to calculate across the zero quark mass from positive to negative mass without any trouble. This is 
because in the deconfining phase the density  of eigenvalues  of the massless Dirac-Wilson operator decreases toward zero (modification\cite{DelDebbio:2010ze}\cite{Patella:2012da} of the Banks and Cacher relation\cite{ref:BanksCasher}.)
 We used this fact to identify the first order chiral phase transition for $N_f=3$ and $6$, which we call ``on the $K_c$ method''\cite{iwa96}, and to find the fact that for $N_f \ge 7$ there is no confining phase at the massless quark in the strong coupling limit.

Thirdly, because the chiral symmetry is explicitly broken in the Wilson action, we have to tune the hopping parameter $K$.
The quark mass does have $1/N$ correction. 
The mass at $N=12$ differ from that at $N=16$ with order of $O(0.001)$, while that at $N=8$ with order of $O(0.005)$ in our case.
We estimate the effect on the meson propagator of this difference. The effect by the difference of $O(0.001)$ is one of order smaller than the statistical errors and that of $O(0.005)$ is order of a half of one standard deviation.
In total, we estimate the smallness of the difference is enough for the accuracy we take in this article. 
 
Now, let us show the results, starting with the $N_f=16$ case.
As mentioned earlier, within the two-loop perturbation,
the IR fixed point is 
$ \beta^*=11.48,$  which is RG scheme independent.
On the other hand, the coupling constants in different RG schemes are related to each other by a constant as $\beta_1=\beta_2+c_{12}$
in the one-loop approximation.
For example\cite{ratio scale parameter},
the lattice coupling constants $\beta_{\mathrm{RG}}$ and $\beta_{\mathrm{one-plaquette}}$ for one-plaquette action are related to that in the continuum theory $\beta_{\overline{\mathrm{MS}}}$ (in the modified minimal subtraction scheme) as
$$\beta_{\mathrm{RG}}=\beta_{\overline{\mathrm{MS}}}-0.3$$
and 
$$ \beta_{\mathrm{one-plaquette}}=\beta_{\overline{\mathrm{MS}}}+3.1.$$
It is well-known that the convergence of the perturbation by the $g_{\mathrm{one-plaquette}}$ is poor in general.
The contribution of higher order terms will be large.
On the other hand,
the lattice coupling constant $\beta_{\mathrm{RG}}$ is close to  $\beta_{\overline{\mathrm{MS}}}$\cite{shrock}
and therefore we may expect that the higher-order contribution is not so large and the location of the fixed point is close to 11.2 in  
$\beta_{\mathrm{RG}}$ from the two-loop estimate and the above relation.

\begin{figure}[thb]
\begin{center}
		\includegraphics [width=7.5cm]{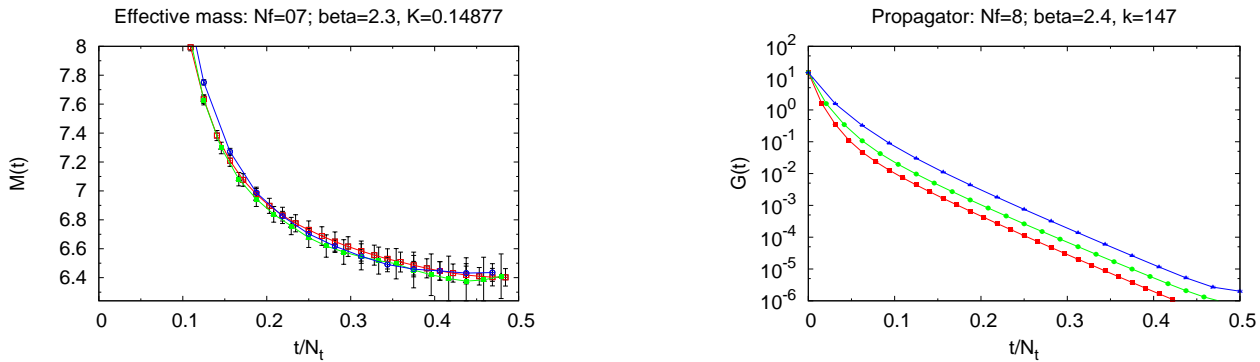}
\vspace{1cm}
\caption{Scaled effective mass plots for $N_f= 7$ at $\beta=2.3$: 
three sets of symbols are $N=16$ (red square), $N=12$ (green circle) and $N = 8$ (blue triangle).}
\label{effective mass plot; 7}
\end{center}
\end{figure}

\begin{figure}[htb]
\begin{center}
	\includegraphics [width=7.5cm]{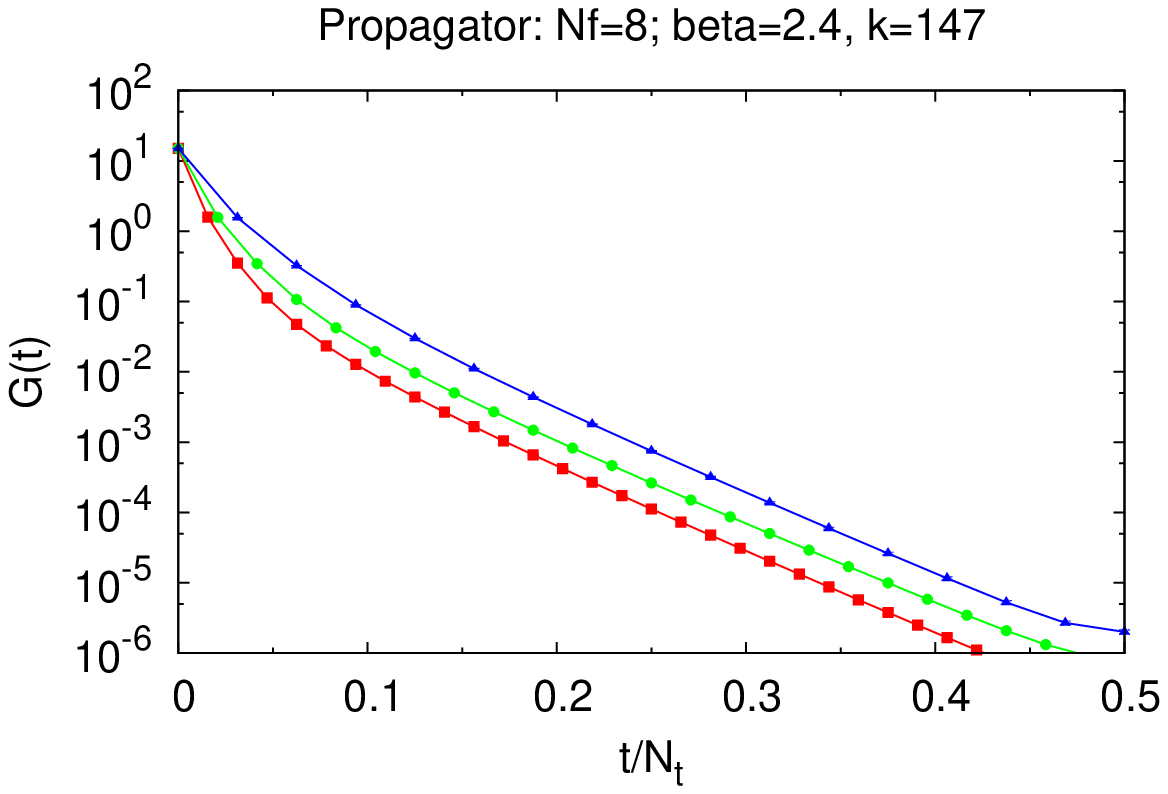}
\vspace{1cm}
\caption{Propagators for $N_f=8$ at $\beta=2.4$:
three sets of symbols are $N=16$ (red square), $N=12$ (green circle) and $N = 8$ (blue triangle).}
\label{propagator}
\end{center}
\end{figure}


In order to find the fixed point from our proposal,we make several trials including those at $\beta=10.0, 10.5, 11.0,$ and $11.5$. We find the three sets of data and the lines connecting them are apparently different from each other at $\beta=11.0$ (Fig. 1; left panel) and they approach closer by decreasing $\beta$ as $11.0, 10.5.$  On the other hand, at $\beta=10.0$ (Fig. 1; right panel) they are apart each other again but they approach closer by increasing $\beta$ as $10.0, 10.5.$  This suggests that there is an IR fixed point between $\beta = 10.0$ and $11.0.$ We indeed find, as shown in  Fig. 2, that the three sets of the scaled effective mass plots are almost degenerate at $\beta=10.5$ and $K=0.1292.$ We see that three lines almost overlap for $\tau \ge0.1$. Only in the small $\tau$ region ($\tau\le 0.1$) we see the differences. 
We interpret the difference for $\tau \le 0.1$ is due to the fact that $N$ is not large enough to remove the effect of the UV cutoff $\mu=a^{-1}.$

The fact that our method identifies the location of the IR fixed point at  a value expected from the perturbation theory, together with the fact that three lines almost overlap, strengthens our confidence in the validity of our approach.

We make similar process for $N_f=12, 8$ and $7$ as the $N_f=16$ case. In Figs. 3 and 4 are shown the results.
The qualitative feature of our results are the same. If we choose a very particular $\beta$ for each $N_f$, 
the data and three lines almost overlap for $\tau \ge0.1,$ as shown in the Figures.
In the small $t$ region ($t/N_t\le 0.1$) we find the differences.  Since they are similar to the case of $N_f=16$, we do not present them here.  


Finally 
we identify the IR fixed points at $\beta^*=10.5\pm 0.5$ for $N_f=16$;  $3.0\pm 0.2$ for $N_f=12$;  $2.4\pm 0.1$ for $N_f=8$; and $2.3\pm 0.05$ for $N_f=7.$ 

On the other hand, in the $N_f=6$ case, there is a chiral phase transition point at finite $\beta$ when $N$ is finite\cite{iwa96}.
If we would perform a program similar to the above (by fixing $\beta$ and increasing the lattice size $N$), then at some $N$ the system would end up with the confining phase rather than the chiral symmetric phase (to which the conformal fixed points belong). 
Thus the IR behavior would be completely different. It cannot be a conformal field theory.

Thus our results at the finite lattice size (up to $16^3 \times 64$) are consistent with that the conformal window is $7 \le N_f  \le 16$.
However we do not exclude  the possibility of the ``walking scenario" that the RG beta function is anomalously small near the edge of the conformal window (e.g. $N_f=7$ or $8$), and for a larger $N$  an undiscovered chiral phase transition point happens to appear at some value of $\beta$ and  the chiral phase transition eventually occurs in the infinite $N$ limit. 

Eq.(\ref{beta_function}) may be used to verify whether the sign of the beta function changes at the fixed points identified, calculating the beta function just above and below the critical coupling constant. The formula contains the differences of the scaled effective masses in the denominator and numerator. The differences are of order $0(1)\%$  of the scaled effective masses and the errors of our data are also of the same order. Therefore it is hard to estimate numerically the beta function just above and below the critical coupling constant with our present data. 

The simulation at larger lattices will clarify which scenario is realized in the continuum limit,  verifying first whether the fixed point does stay at the identified fixed point and second whether the sign change occurs.

\begin{figure*}[htb]
\begin{center}
	\includegraphics [width=7.5cm]{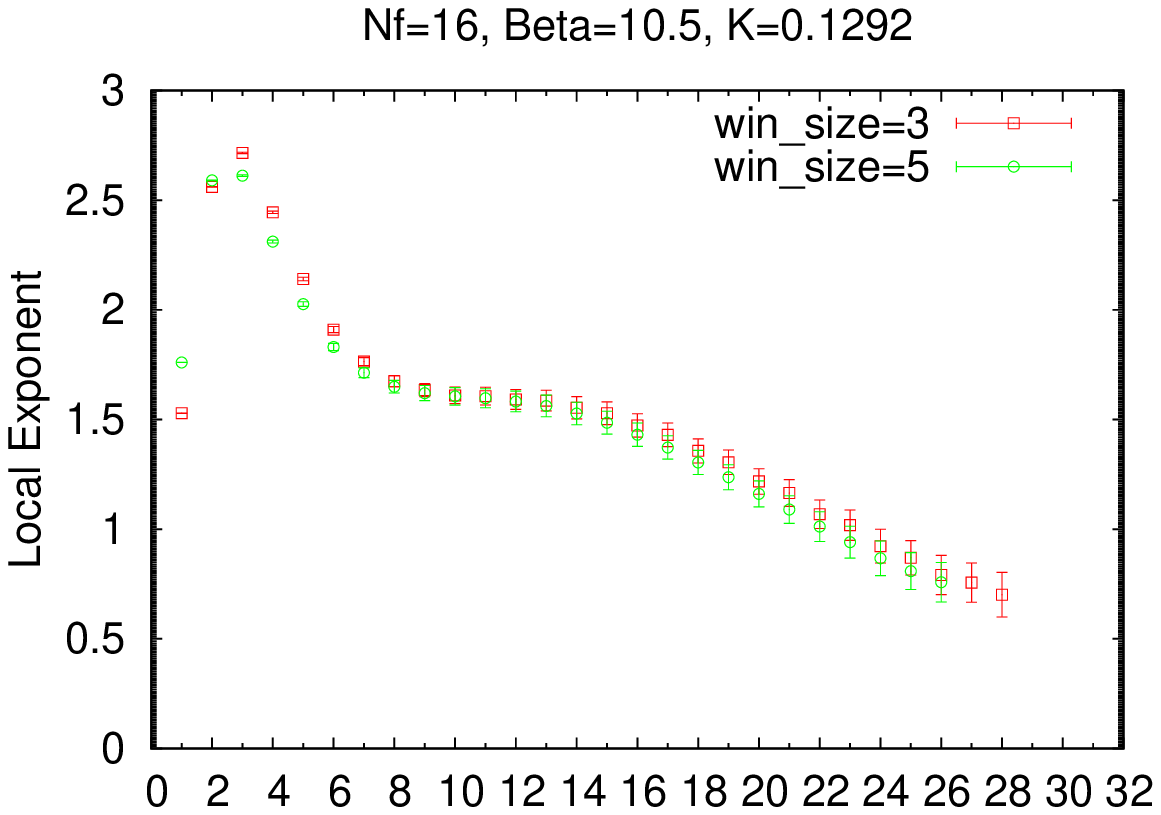}
\hspace{1cm}
	\includegraphics [width=7.5cm]{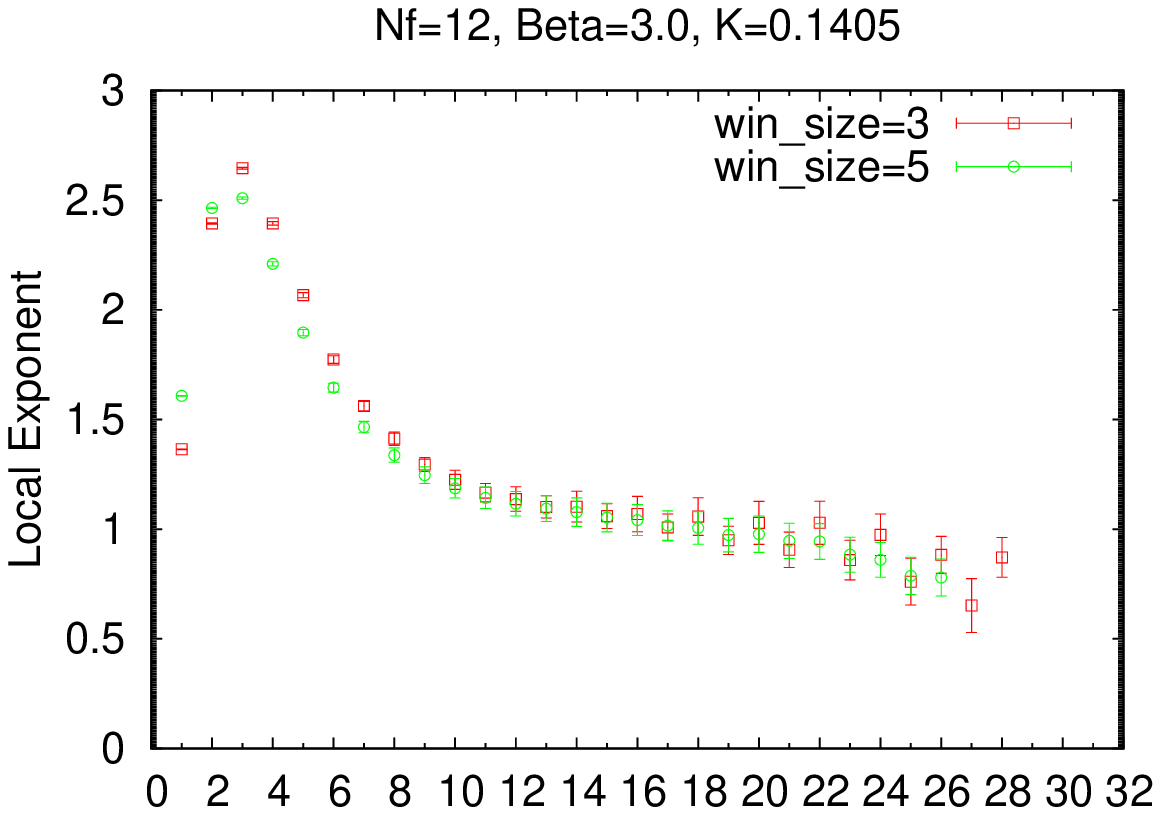}
\vspace{1cm}
		\caption{Local exponent $\alpha(t)$: $N_f=16$ in the left panel; $N_f=12$ in the right panel}
\label{local exponent 12 and 16}
\end{center}
\end{figure*}

\begin{figure*}[htb]
\begin{center}
\includegraphics [width=7.5cm]{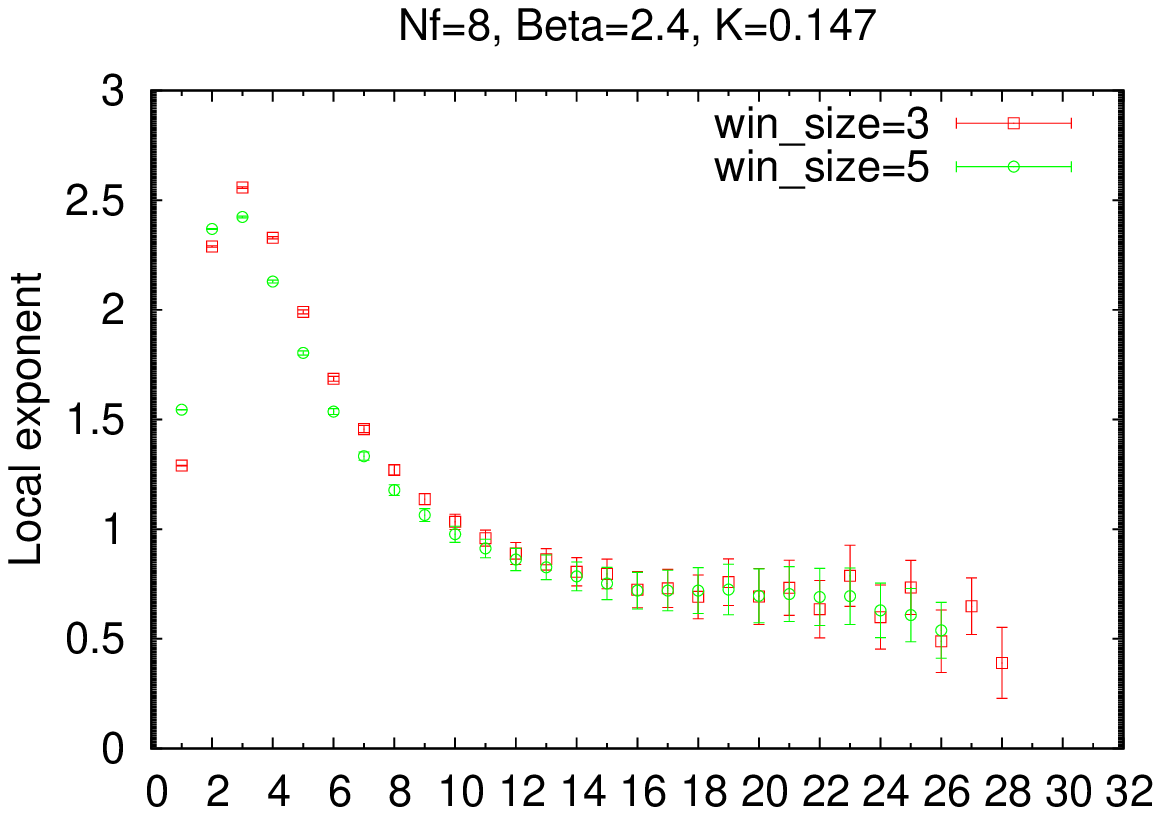}
\hspace{1cm}
	\includegraphics [width=7.5cm]{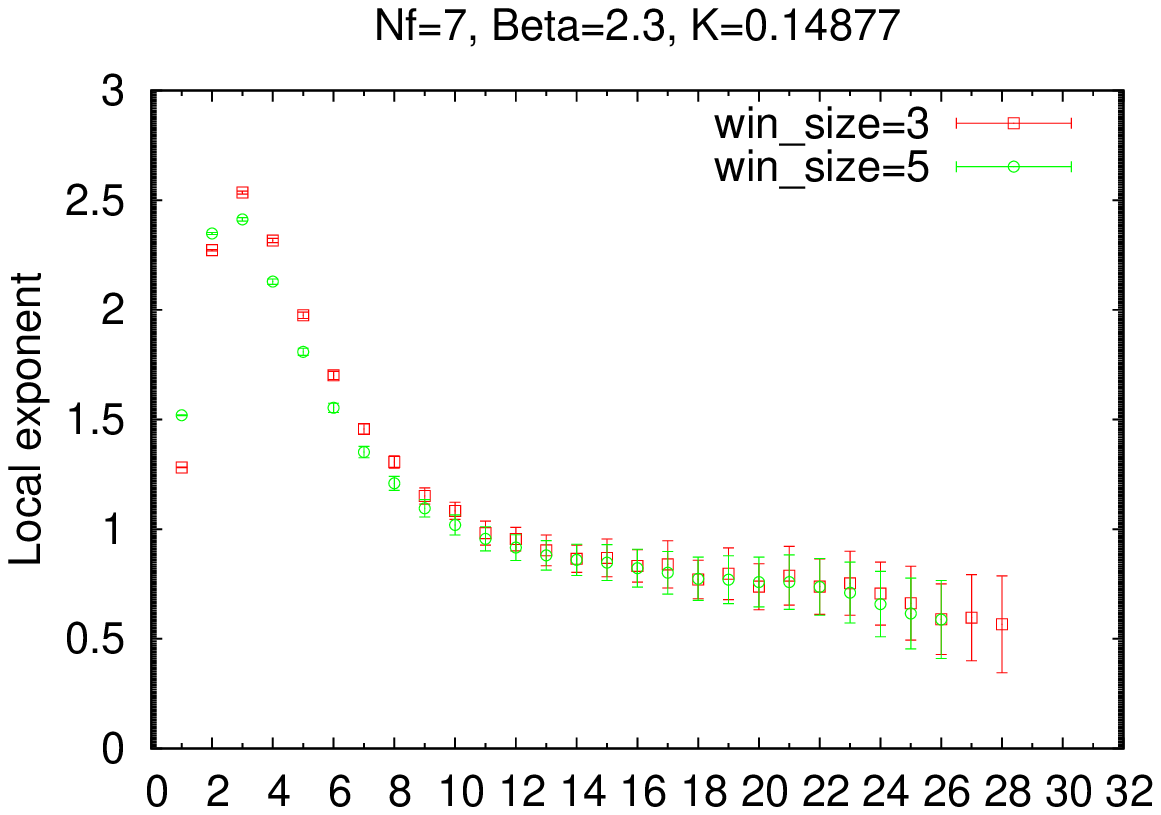}
\vspace{1cm}
\caption{Local exponent $\alpha(t)$: $N_f=8$ in the left panel; $N_f=7$ in the right panel}
\label{local exponent 7 and 8}
\end{center}
\end{figure*}


We note that it seems possible to extract the mass anomalous dimension $\gamma^{*}$ using the scaling of propagators Eq.(\ref{mass dimension})
in the continuum limit  $N  \rightarrow \infty$.
In Fig.\ref{propagator} we show the results for the propagator on the three lattice sizes in the $N_f=8$ case.
The data are depicted on a logarithmic scale.
We see the data roughly scale at  $\tau \ge 0.1$.
Similar results are obtained also for $N_f=16, 12$ and $N_f=7.$

We try to estimate  $\gamma^{*}$  using the scaling law in Eq.\ref{mass dimension}.
For each $N_f$ there are three sets for $N$ and $N'$: (8,16), (8, 12), (12,16).
In each set we estimate $3.0 - 2 \gamma^{*} $ and statistical errors at each corresponding $\tau$, treating data at each $\tau$ as statistically independent.
The statistical errors are small at small $\tau$ and become large at large $\tau$, so we quote the results at $\tau=0.125.$
The results are in order of (8,16), (8, 12), (12,16):
\begin{align*}
N_f=16:&  -0.0309(24), -0.0516(24), -0.0017(64). \cr 
N_f=12: & 0.0337(35), 0.0086(55), 0.0692(105). \cr
N_f=8:  & 0.15148(51), 0.1047(77), 0.1289(125). \cr
N_f=7:	& 0.1767(55), 0.1652(83), 0.1929(163).
\end{align*}
In the case $N_f=16$, the perturbative computation gives $\gamma^*= 0.026$. The order of magnitude for $N_f=16$ is reasonable given the large statistical error.
The order of estimated anomalous mass dimensions for $N_f=12$ is apparently smaller than those reported in the literature (See for example \cite{Hasenfratz2013}).
The reason may be $N$ and $N'$ are not large enough to obtain anomalous mass dimensions in the continuum limit.

To gain an intuition about the systematic error, we can perform the same scaling analysis for the free massless Wilson fermion. It shows the ``anomalous dimension" of $-0.06, -0.07,-0.04$ at $\tau = 0.125$ (for the same lattice size ratio above) while the scaling of effective masses is less affected by the finite size effects. In general, the effective masses scale better than the amplitude even with smaller $N$. To find the complete resolution of the issue we have to simulate with lager $N$. We leave it for the future study.



Let us finally discuss the physical meaning of the shape of the effective mass plot at the fixed point.
In order to investigate the dynamics of the theory,
we proposed a detailed analysis of temporal propagators 
which we call the ``local-analysis" of propagators~\cite{coll2014}.
We parametrize the propagator $G(t)$ as
\begin{equation} 
G(t) = c(t)\,  \frac {\exp(-m(t)\, t)}{t^{\, \alpha(t)}},
\label{local}
\end{equation}
It is possible to determine $c(t_0), m(t_0), \alpha(t_0)$ locally, using three point data
$G(t_0 ), G(t_0+1), G(t_0 +2).$ This is not a fit, but a parametrization.
In addition to this parametrization, we also perform a fit using data at five points and check  the fit and the parametrization are in good agreement, which implies that the parametrization represents the characteristics of the propagator reasonably well.

The results of the local parameters are similar to those obtained in our previous article\cite{coll2014} 
except for some quantitative differences.
Our current data actually suggests that the data in our previous work\cite{coll2014} are slightly off the IR fixed points.
Nevertheless almost  all what were discussed there can be applied also here.
To avoid repetition we do not plot $m(t)$, but plot $\alpha(t)$ for $N_f=16, 12, 8,$ and $7$ in Figs. \ref{local exponent 12 and 16} and \ref{local exponent 7 and 8} since the $\alpha(t)$ mostly represents the characteristics. We observe an interesting change of the form of $\alpha(t)$ from $N_f=16$ to $N_f=7.$ For $N_f=16$ there is a shoulder around  $\alpha \sim1.5$ at $8\le t \le 14.$ On the other hand for $N_f=8$ and $7,$ there is a plateau around $\alpha \sim 0.7$ at $t \ge 14.$ The $N_f=12$ case shows a transition between them.

\begin{figure}[htb]
  \begin{center}
	\includegraphics [width=7.5cm]{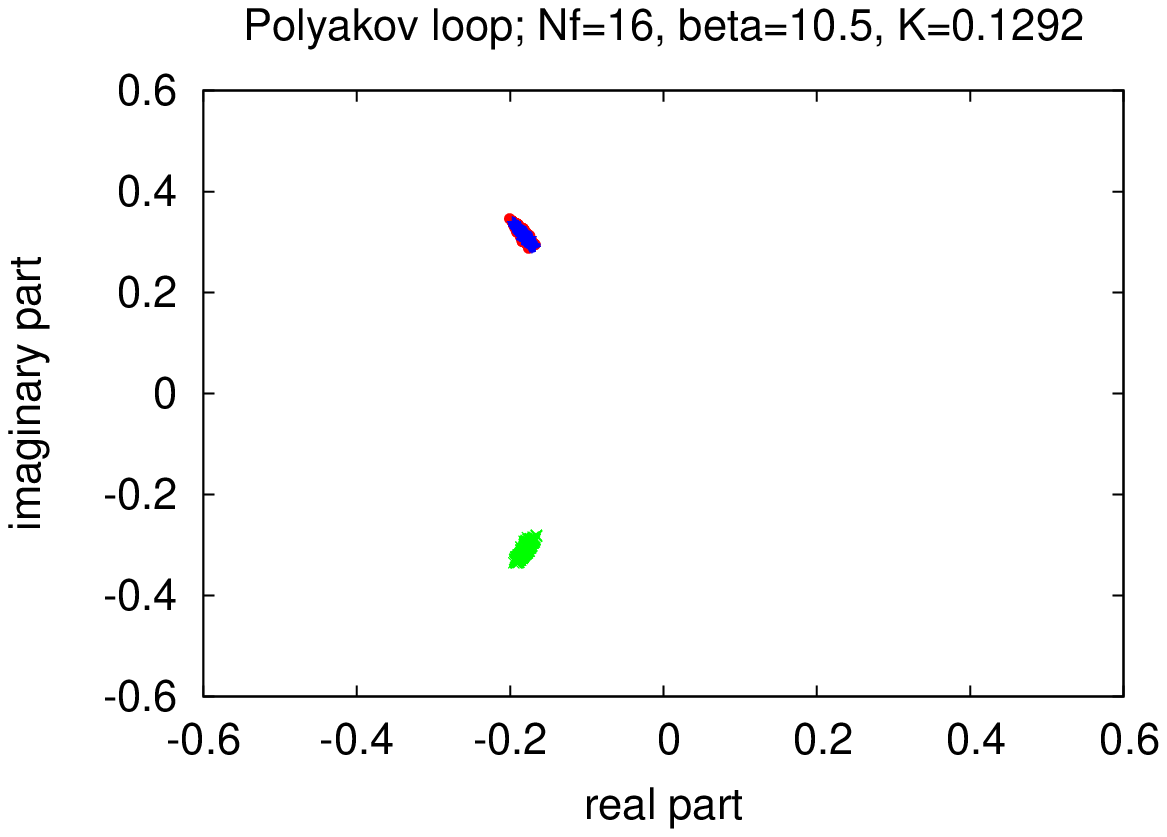}
\vspace{1cm}
\caption{Scattered plot of Polyakov loops for $Nf=16$ in $(x , y, z)$ directions; (red, green, blue), respectively.
}
\label{polyakov loop}
  \end{center}
\end{figure}

In addition to the local analysis of propagators we examine the Polyakov loops in spatial directions to investigate the vacuum structure.
The results are also similar to those in Ref.\cite{coll2014}.
As a typical example we show the scattered plot of the Polyakov loops for$N_f=16$ in Fig.\ref{polyakov loop}.

From these results  we may conclude that the vacuum of the $N_f=16$ is the $Z(3)$ twisted one modified by non-perturbative effects and the meson is an almost free fermion state in the twisted vacuum. The $t$ dependence of $\alpha(t)$ is very similar to that of free fermions in the $Z(3)$ twisted vacuum.  See Ref.\cite{coll2014}  for the details.

The plateaus at $14 \le t \le 24$ in the $N_f=8$ and $7$ cases,
taking values $\alpha(t)  \sim 0.7 \pm0.1$ is well described by an unparticle meson model and
 $\gamma^*\sim 1.3\pm 0.1$ from the formula $2-\alpha =\gamma^*$\cite{coll2014}.
 We need the data with high statistics for a more precise value of $\gamma^*.$

 
We would like to stress that a conformal field theory is completely different from QCD in the point that there is no
dimensional parameter such as $\Lambda_{\mathrm{QCD}}$.
In QCD if $N a$ is large enough compared with $\Lambda_{\mathrm{QCD}}$, boundary effects can be neglected and it can be  assumed the limit $N=\infty$ is taken.
However the boundary effects are essential even at any large lattice $N$ in the conformal field theories because there is no other natural scale to compare.
Note that the propagators Eqs. (7) and (11) are functions of the scaled time $\tau$ which takes value $0.0 \le \tau \le 1.0.$ Clearly the function depends on the boundary condition as well as the aspect ratio even if we take $N\to \infty$ limit. 
Of course, to be clear, this does not mean that the {\it local} physics of the conformal field theory  depends on the boundary conditions we use. 
We note the zero momentum propagator in our definition \eqref{propagator} may not be a local variable because we have summed over spatial coordinates before taking the continuum limit.





In the near future we would like to perform the program with larger lattice sizes and more statistics to derive 
the anomalous mass dimension using  Eq.\ref{mass dimension}, and the relation of the eigenvalue density of the Dirac-Wilson operator and the anomalous index\cite{DelDebbio:2010ze}\cite{Patella:2012da}\cite{Hasenfratz2013}. It would be intriguing to compare them with the value from the unparticle meson model.

The calculations were performed with HA-PACS computer at CCS, University of Tsukuba and SR16000
at KEK. We would like to thank members of CCS and KEK for their strong support for this work.


\appendix

\end{document}